\pdfoutput=1

\documentclass[referee]{raa}          

\usepackage{graphicx,times}             
\usepackage{natbib}
\usepackage{amssymb,amsmath}
\bibpunct{(}{)}{;}{a}{}{,}
\usepackage[a4paper, left=2.5cm, right=2.5cm, top=2.5cm, bottom=3.0cm]{geometry}
\usepackage[pagebackref=true]{hyperref}

\begin{document}

  \title{On the performances of estimating stellar atmospheric parameters from CSST broad-band photometry}

   \volnopage{Vol.0 (20xx) No.0, 000--000}      
   \setcounter{page}{1}         

   \author{Rui-Feng Shi
      \inst{1,2}
   \and Yang Huang
      \inst{1,3,7}
   \and Xin-Yi Li
      \inst{4}
    \and Hua-Wei Zhang
      \inst{5,6}
   }

   \institute{School of Astronomy and Space Science, University of Chinese Academy of Sciences, Beijing 100049, China;\\
        \and
             South-Western Institute for Astronomy Research, Yunnan University, Kunming 650500, China;\\
        \and
             CAS Key Lab of Optical Astronomy, Key Laboratory of Space Astronomy and Technology, National Astronomical Observatories, Chinese Academy of Sciences, Beijing 100012, China;\\
        \and College of Physics and Electronic Engineering, Qilu Normal University, Jinan 250200, China;\\
        \and Department of Astronomy, School of Physics, Peking University, Beijing 100871, China;                         \\
        \and Kavli Institute for Astronomy and Astrophysics, Peking University, Beijing 100871, China;      \\
        \and Corresponding author: Yang Huang ({\it huangyang@ucas.ac.cn\rm)}\\
\vs\no
   {\small Received 20xx month day; accepted 20xx month day}}

\abstract{ Deriving atmospheric parameters of a large sample of stars is of vital importance to understand the formation and evolution of the Milky Way. Photometric surveys, especially those with near-ultraviolet filters, can offer accurate measurements of stellar parameters, with the precision comparable to that from low/medium resolution spectroscopy. In this study, we explore the capability of measuring stellar atmospheric parameters from CSST broad-band photometry (particularly the near-ultraviolet bands), based on synthetic colors derived from model spectra. We find that colors from the optical and near-ultraviolet filter systems adopted by CSST show significant sensitivities to the stellar atmospheric parameters, especially the metallicity. According to our mock data tests, the precision of the photometric metallicity is quite high, with typical values of 0.17 dex and 0.20 dex for dwarf and giant stars, respectively. The precision of the effective temperature estimated from broad-band colors are within 50\,K. 
}

   \authorrunning{R.-F. Shi, Y. Huang, X.-Y. Li \& H.-W. Zhang}       
   \titlerunning{Estimating stellar atmospheric parameters for CSST}  

   \maketitle

\section{Introduction}           

The current oldest stars in the Milky Way are supposed to be the Population II (Pop II) stars whose progenitors are the first stars--the so called Population III (Pop III) stars (\citealt{Lardo+etal+2021}). It is thought that the first stars formed in hundreds of millions of years just after the Big Bang as a consequence of condensation of cosmological mini-halos. Consisting of H, He and trace amounts of Li, mini-halos are cooled via molecular hydrogen with cooling efficiency far less than that of metal. Therefore, the first stars are supposed to be massive stars (\citealt{Ishigaki+etal+2021, Zepeda+etal+2022}). Metals formed in their short lives are ejected into interstellar medium through supernova explosions, which is the first time for the enrichment of primordial gas that leads to the formation of the second stars with low masses (\citealt{Umeda+Nomoto+2002, Heger+Woosley+2002, Heger+Woosley+2010, Limongi+Chieffi+2012, Nomoto+etal+2013, Ishigaki+etal+2018}).

Metallicity is the fossil record of the birth place of the star. We thus can figure out the nature of the first stars by searching for the most metallicity deficient stars in the Milky Way. As an example, we can constrain the baryon number to photon number ratio $\eta$ (\citealt{Beers+Christlieb+2005}) by measuring abundance of Li in very metal-poor stars. Moreover, we can determine metallicity distribution function (MDF) of the Galaxy to explore the chemical enrichment at the early stage of the Milky Way (\citealt{Bonifacio+etal+2021}).

The key to the above scientific problems is the precise estimate of stellar atmospheric parameters for a huge sample of stars, in particular the metallicity. These have partly been done by spectroscopic surveys in the past few decades, which greatly improves our understanding of the assembly history of the Galaxy. However, the raw number of stars targeted by spectroscopic surveys is limited to the order of $10^{6}$, far behind the number achieved by photometric survey (like Pan-STARRS1; \citealt{Chambers+etal+2016}) and astrometric one (like Gaia; \citealt{Brown+etal+2016}). Moreover, the complex selection functions should also be properly dealed with in the studies based on spectroscopic data (\citealt{Ivezic+etal+2008, Huang+etal+2022}). On the other hand, as summarized in Huang et al. (2022), stellar atmospheric parameters in particular metallicity of dozens of million stars can be precisely measured from phtometric surveys with near-ultraviolet bands such as the Sloan Digital Sky Survey (SDSS, \citealt{York+etal+2000}), the SkyMapper Southern Survey (SMSS, \citealt{Wolf+etal+2018, Onken+etal+2019}), the Pristine survey (\citealt{Starkenburg+etal+2017}), the Stellar Abundance and Galactic Evolution survey (SAGES, \citealt{Zheng+etal+2018, Fan+etal+2023}), the Javalambre Physics of the Accelerating Universe Astrophysical Survey (J-PAS, \citealt{Benitez+etal+2014}), the Javalambre Photometric Local Universe Survey (J-PLUS, \citealt{Cenarro+etal+2019}), the Southern Photometric Local Universe Survey (S-PLUS, \citealt{Mendes de Oliverira+etal+2019, Whitten+etal+2021}) and the Chinese Space Station Telescope (CSST, \citealt{Zhan+2011}). The recent efforts based on surveys with narrow/medium near-ultraviolet bands have shown the precision of metallicity is comparable to that from low/medium resolution spectroscopy (e.g., \citealt{Huang+etal+2019, Huang+etal+2022, Huang+etal+2023, Lin+etal+2022})

\begin{figure}
   \centering
  \includegraphics[width=14cm, angle=0]{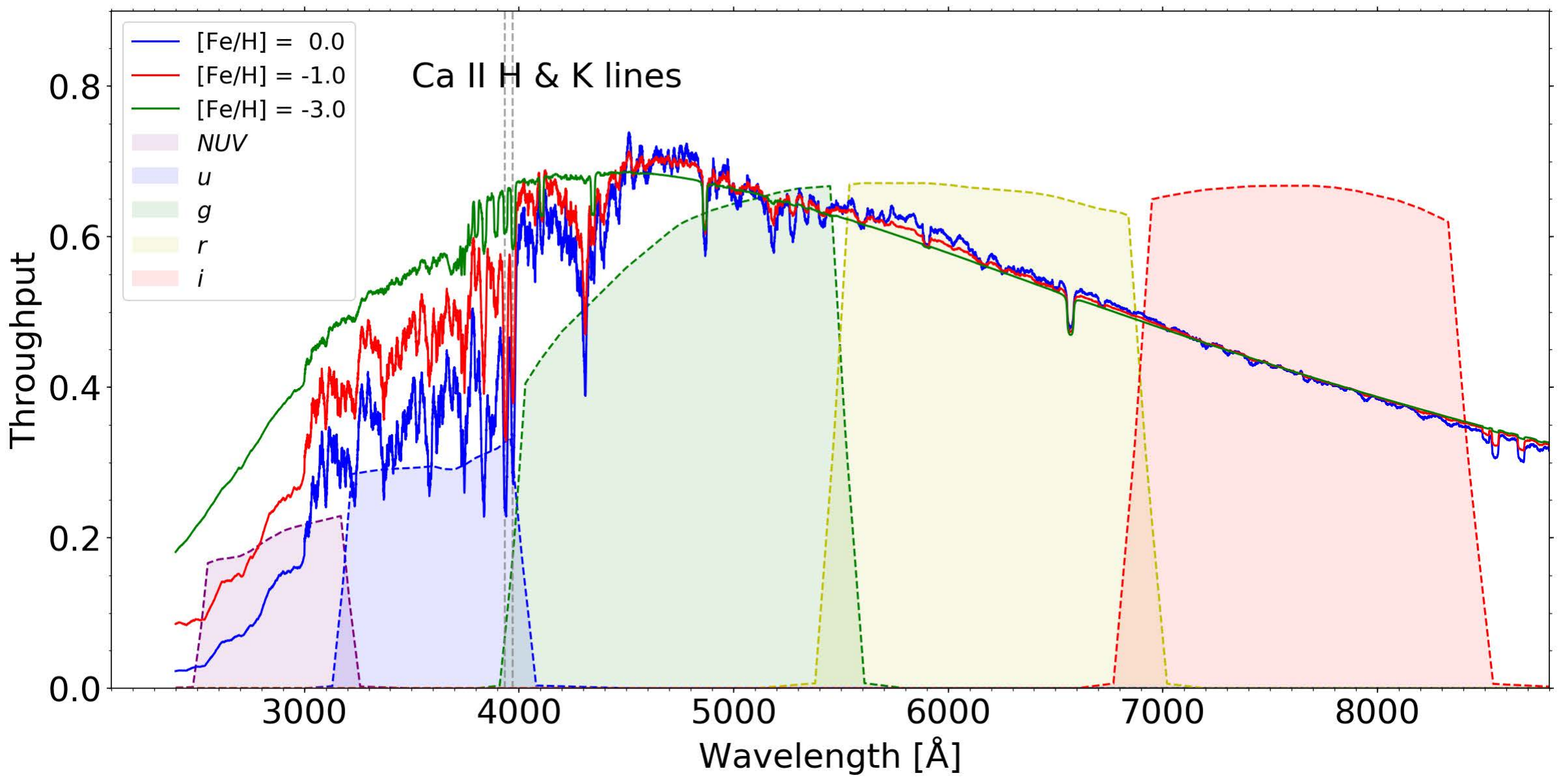}
   \caption{The sensitivity of CSST broad-band filters to stellar metallicity. Three spectra with various metallicities are shown, with the blue, red and green one represents [Fe/H] equal to 0, $-$1, $-$3, respectively. They are normalized using fluxes ranging from 7,000~\AA\ to 7,500~\AA. The effective temperature and surface gravity of them are the same, with values of 5,800 K and 4.50, respectively.} 
   \label{Fig1}
   \end{figure}

In this study, we focus on the CSST, a 2 m space telescope with a field of view 1.1 $\rm deg^{2}$. It is planned to be launched into the low earth orbit (LEO) in 2025 to carry out large-scale sky survey covering nearly 17,500 $\rm deg^{2}$ at high Galactic latitude ($b \ge 15^{\circ}$) with $g$ band limiting magnitude down to around 26.3 (5$\sigma$ point-like source). The CSST is equipped with seven filters $NUV$, $u$, $g$, $r$, $i$, $z$, $y$ with wavelength ranging from 2,000 \AA\, to 1.1$\rm \mu$m. With such large sky coverage and survey depth, the future CSST data will provide revolutionary view of our Galaxy, especially the outer halo in an unprecedented volume. It should be specially mentioned that the atmospheric parameters of a large sample of stars will be achieved with its $NUV$ and $u$ bands, as well as slitless spectra although with shallower limiting magnitude. This study attempts to explore the capability of measuring stellar atmospheric parameters from the CSST broad-band colors. As a glance in Figs.\,1 and 2, the CSST $NUV$ and $u$ bands show significant sensitivities to the stellar surface gravity and metallicity.

This paper is organized as follows. In Section 2, we briefly introduce the adopted theoretical spectra and calculations of stellar colors. In Section 3, the sensitivity of color to stellar atmospheric parameters is explored in details. Classification of stars is done, as illustrated in detail in Section 4. Tests on estimates of metallicity and effective temperature from the CSST filter systems are shown in Section 5 and Section 6, respectively. Finally, we summarize our results and make discussions in Section 7.

\begin{figure}
   \centering
  \includegraphics[width=14cm, angle=0]{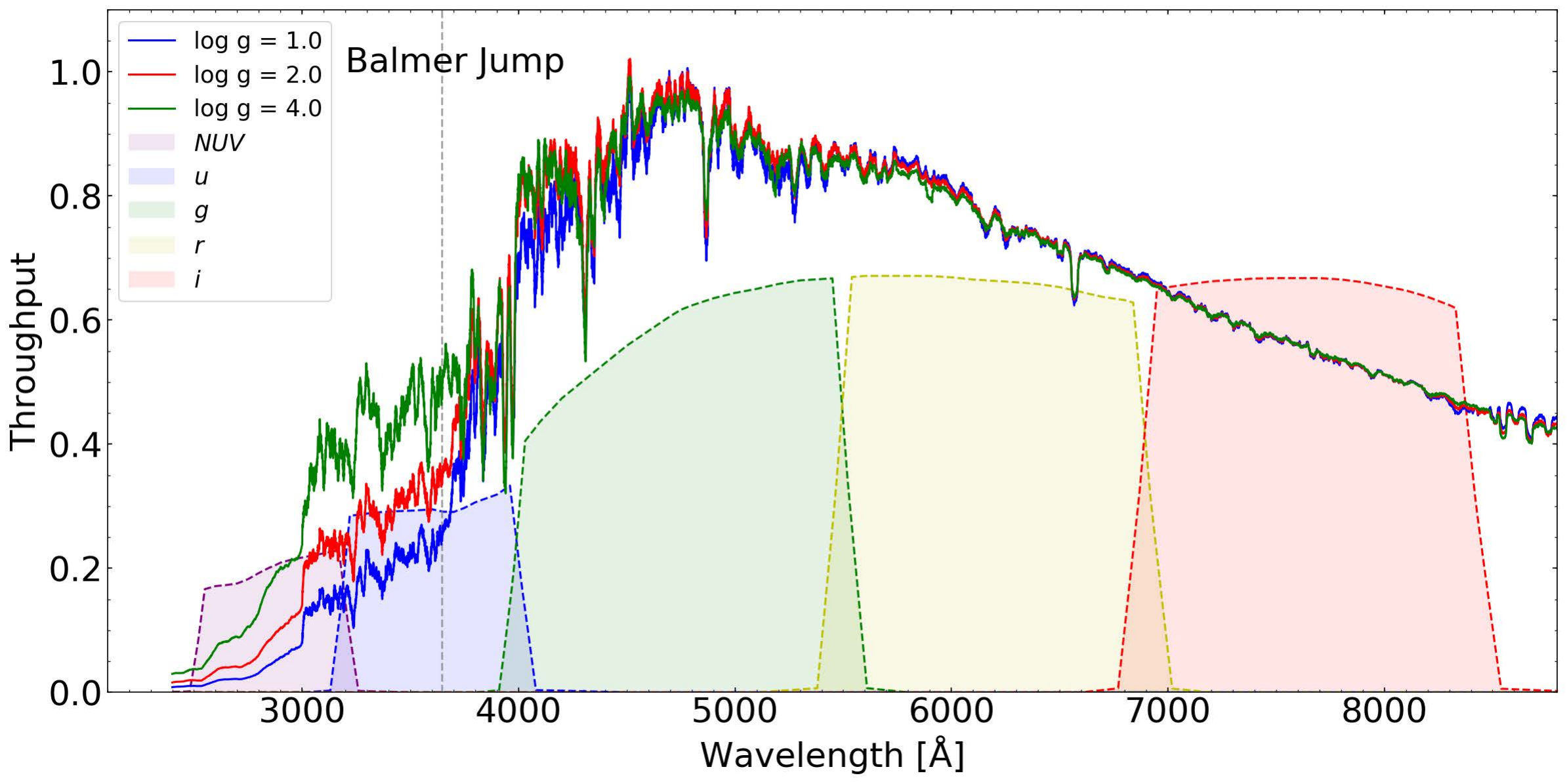}
   \caption{The sensitivity of CSST broad-band filters to surface gravity. Three spectra with various surface gravities are shown. They are again normalized using fluxes ranging from 7,000~\AA\ to 7,500~\AA. The effective temperature and metallicity of them are the same, with values of 5,800 K and 0.0, respectively.} 
   \label{Fig2}
   \end{figure}

\section{Theoretical spectra and calculations of stellar colors}
\subsection{Theoretical Spectra}
Theoretical spectra library can offer us high-resolution spectra with wide parameter space coverage, the high-resolution spectra can be degraded to any lower one upon our requests. For this study, we have downloaded 4,542 high resolution spectra from the PHOENIX synthetic library\footnote{https://phoenix.astro.physik.uni-goettingen.de/} (\citealt{Husser+etal+2013}). It covers 3,000 K $\le T_{\rm eff} \le$ 10,000 K, 0.0 $\le$ log $g \le$ 6.0, $-$4.0 $\le \rm [Fe/H] \le$ +0.5, and 0.0 $\le \rm [\alpha/Fe] \le$ 0.6; the parameter coverages are shown in Fig.\,3. The high resolution spectra cover the wavelength range from 500 to 55,000 ~\AA\ with resolving power of $R$ = 500,000 in the optical and near infrared (3,000 to 25,000 ~\AA), $R$ = 100,000 in the infrared region (25,000 to 55,000 ~\AA), and $\Delta\lambda$ = 0.1 ~\AA\ in the UV (500 to 3,000 ~\AA).

\begin{figure}
   \centering
  \includegraphics[width=14cm, angle=0]{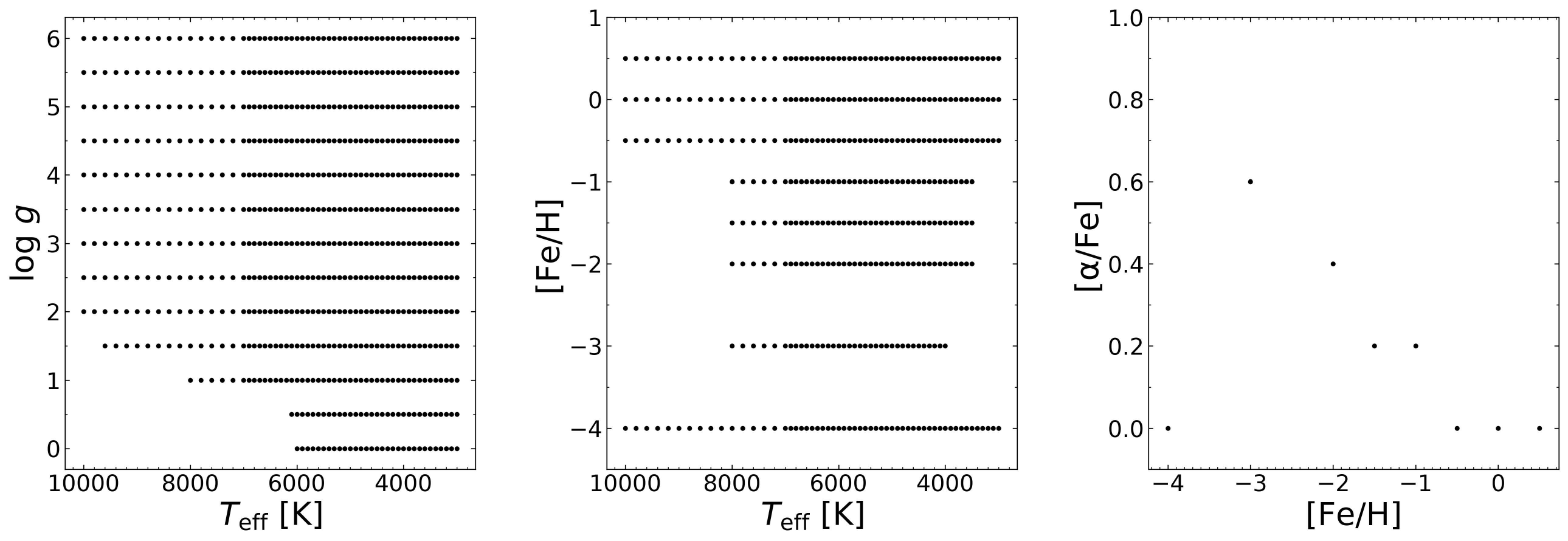}
   \caption{The parameter coverage of the theoretical spectra in $T_{\rm eff}$--log $g$ diagram (left panel), $T_{\rm eff}$--[Fe/H] diagram (middle panel) and [Fe/H]--$\rm[\alpha/Fe]$ plane (right panel). It is worth noting that no alpha-enhanced spectra are provided for ultra metal-poor stars ([Fe/H] = -4.0) in PHOENIX synthetic library.} 
   \label{Fig3}
   \end{figure}

\subsection{Calculations of Synthetic Colors}
Once the transmission curves of the filter systems (see Figs.\,1 and 2) and theoretical spectra are given, synthetic colors can be calculated by proper convolutions (\citealt{Casagrande+VandenBerg+2014}). Here we adopted the AB magnitude system which is defined as:

\begin{equation}\label{eq1}
  m_{\rm AB} = m_{\rm ST} - 5\rm log\lambda_{PIVOT,\zeta} +18.6921, 
\end{equation}
  where m$_{\rm ST}$ is ST monochromatic magnitudes defined as
\begin{equation}\label{eq3}
  m_{\rm ST}=-2.5 \rm log\frac{{\int_{\lambda_{i}}^{\lambda_{f}}\lambda \emph {f}_{\lambda}T_{\zeta}d\lambda}}{{\int_{\lambda_{i}}^{\lambda_{f}}\lambda T_{\zeta}d\lambda}}-21.10,
\end{equation}
  where $\zeta$ is bandpass ranging from $\lambda_{i}$ to $\lambda_{f}$, $ f_{\rm \lambda}$ is the flux at specific wavelength, $\rm T_{\zeta}$ is throughput of a certain filter. $\rm \lambda_{PIVOT,\zeta}$ is defined as
  
\begin{equation}\label{eq2}
  \rm \lambda_{PIVOT,\zeta} =(\frac{\int\lambda T_{\zeta}d\lambda}{\int \frac{T_{\zeta}}{\lambda}d\lambda})^\frac{1}{2}.
\end{equation}

By adopting the transmission curves of CSST filters, we thus integrate magnitudes and their combined colors using the aforementioned downloaded theoretical spectra.

\section{Sensitivity of color to stellar atmospheric parameters}

Based on the above synthetic color calculations, we now can evaluate the sensitivities of CSST colors on estimations of stellar atmospheric parameters. First, the stellar effective temperature can be derived from colors with two bands having separation in central wavelength (e.g. $g-i$, $g-z$ or $g-y$). According to previous studies (e.g., \citealt{Ramirez+Melendez+2005, Casagrande+etal+2010, Huang+etal+2015}), the metallicity effect can not be ignored in effective temperature determinations; but weak metallicity dependence could be found for some $T_{\rm eff}$ -- color relations, where the latter is a combination of optical and near-infrared bands. Therefore, colors of $g-i$, $g-z$ and $g-y$ are adopted for effective temperature estimates; the detail relations will be constructed in Section 6.

\begin{figure}
   \centering
  \includegraphics[width=14cm, angle=0]{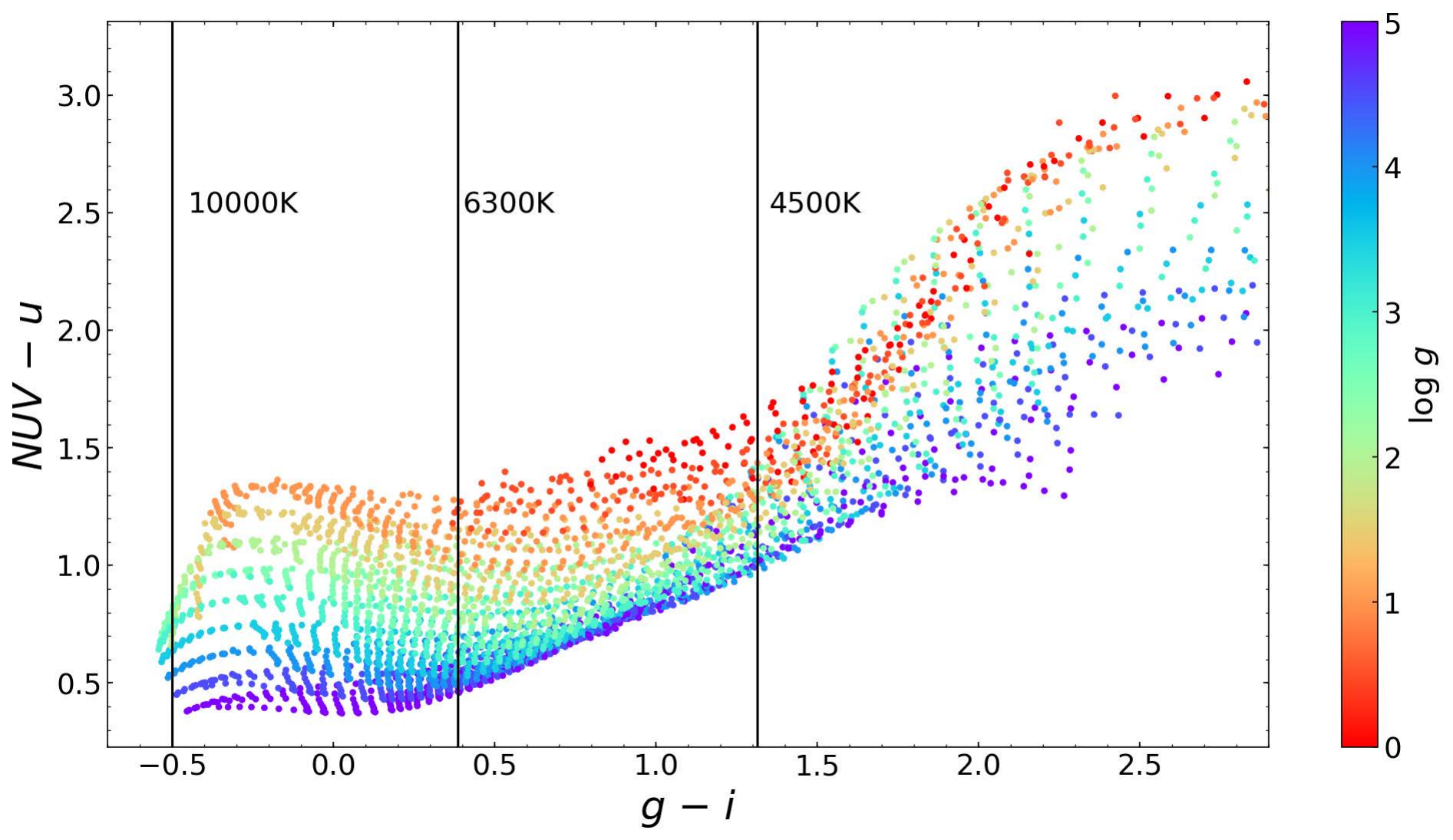}
   \caption{Sensitivity of color to log $g$ of stars from CSST filter systems in the ($NUV-u$) versus ($g-i$) plane, color coded by log $g$, as shown by the right color bars. The values of log $g$ range from 0 dex to 6 dex.}
   \label{Fig4}
   \end{figure}

\begin{table}
\begin{center}
\caption[]{ Sensitivity of color $NUV-u$ to surface gravity.}\label{Tab1}
\begin{tabular}{ccccccc}
  \hline\noalign{\smallskip}
Sensitivity    & 4,500 -- 6,300 K    &    & 6,300 -- 10,000 K    & \\
&    $\mu_{1}$ (mag $\rm dex^{-1}$)    & $\sigma_{1}$ (mag $\rm dex^{-1}$)    & $\mu_{2}$ (mag $\rm dex^{-1}$)    & $\sigma_{2}$ (mag $\rm dex^{-1}$)    & \\
  \hline\noalign{\smallskip}
All stars    & 0.1146    & 0.1004    & 0.1913    & 0.0663 \\ 
1.0 --- 1.5    & 0.2045    & 0.0642    & 0.2409    & 0.0840 \\
1.5 --- 2.0    & 0.2098    & 0.0640    & 0.2005    & 0.0556 \\
2.0 --- 2.5    & 0.1701    & 0.0595    & 0.2222    & 0.0377 \\
2.5 --- 3.0    & 0.1343    & 0.0630    & 0.2356    & 0.0258 \\
3.0 --- 3.5    & 0.0938    & 0.0606    & 0.2135    & 0.0280 \\
3.5 --- 4.0    & 0.0431    & 0.0592    & 0.1895    & 0.0335 \\
4.0 --- 4.5    & 0.0256    & 0.0403    & 0.1582    & 0.0415 \\
4.5 --- 5.0    & $-$0.0007    & 0.0323    & 0.1192    & 0.0531 \\
  \noalign{\smallskip}\hline
\end{tabular}
\end{center}
\tablecomments{1\textwidth}{Here $ \mu_{i}$, $\sigma_{i}$ (i=1, 2) is the mean value and scatter of the sensitivity in various temperature bins, respectively. The first row is the total sensitivity, the remaining rows are for different bins of surface gravity with range specified. The sensitivity of log $g$ from 0 to 1 is not listed since there are no data with log $g$ = 0, 0.5 above 6,000 K. Moreover, stars with log $g > 5.0$ are excluded in the calculation of the sensitivity, given their extremely low possibilities in the real observations.}
\end{table}

Second, color $NUV-u$ is thought as an indicator of surface gravity. As shown in Fig.\,4, this color shows significant sensitivity to log $g$ between 6,300 K and 10,000 K. The sensitivity becomes weak for G/K-type ($T_{\rm eff}$ between 4,500 K and 6,300 K) stars, but basically one can distinguish dwarf and giant stars based on $NUV-u$. Sensitivity is notably reduced for cool stars with $T_{\rm eff}$ \textless \,4,500 K. Quantitatively, the sensitivity is calculated as the gradient of $NUV-u$ along log $g$, $\Delta (NUV - u)$/$\Delta$ log $g$, at a given $g-i$ bin (equivalent to the effective temperature bin). The results are listed in Table 1. The sensitivity from 6,300 K to 10,000 K is quite high with the mean sensitivity of 0.19 mag dex$\rm ^{-1}$, and a small scatter of 0.07 mag dex$\rm ^{-1}$. For G/K-type stars, the sensitivity remains quite well for giant stars and drops toward increasing log $g$. In summary, the color $NUV-u$ is a good indicator to classify dwarf/giant for any star hotter than 4,500 K.

CSST color ($u-g$) can be used to estimate stellar metallicity. Fig.\,5 shows the sensitivity of color to metallicity for the dwarf and giant stars. It can be clearly seen that the color shows significant sensitivity to stellar metallicity spaning a wide temperature range for both dwarf stars (4,500 K to 8,000 K) and giant stars (4,500 K to 6,500 K). Quantitatively, the sensitivity is calculated as the gradient of $u-g$ along [Fe/H], $\Delta (u-g)$/ $\Delta$ [Fe/H], at a given $g-i$ bin. The results are listed in Table 2. Generally, the sensitivity decreases with decreasing [Fe/H] but holds quite well until [Fe/H] down to $-3.0$. Overall, the sensitivity of giant stars is larger than that of dwarf stars.

We caution that the above sensitivity analysis is based on the PHOENIX theoretical library. However, the synthetic spectra cannot model the real ones perfectly, especially in the ultraviolet region (wavelength smaller than 3,000 \AA), which is the key region to derive stellar parameters. This would be a limitation of our sensitivity analysis. For example, the recent study (by \citealt{Lu+etal+2023}) indicates that $NUV$ from {\it GALEX} shows significant sensitivities to metallicity, while the CSST $NUV$ predicted by synthetic spectra only shows moderate sensitivities to metallicity.

\begin{figure}
   \centering
  \includegraphics[width=14cm, angle=0]{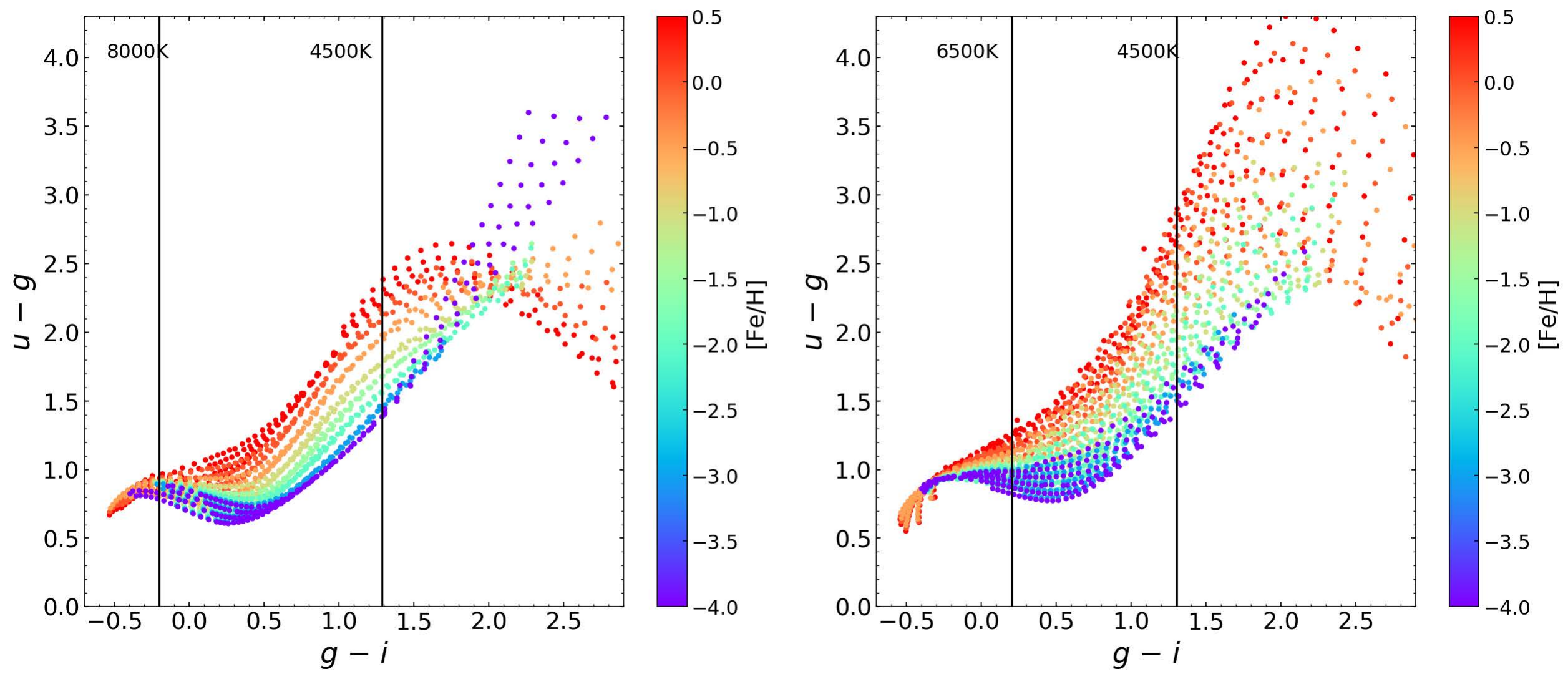}
   \caption{Sensitivy of color to metallicity from CSST filter systems in the ($u$-$g$) versus ($g$-$i$) plane for the dwarf stars (log $g \ge$ 3.5 dex; left panel) and giant stars (log $g < 3.5$ dex; right panel), color coded by metallicity [Fe/H], as shown by the right color bars. The values of [Fe/H] range from $-$4 dex to $+$0.5 dex for both panels.} 
   \label{Fig5}
   \end{figure}

\begin{table}
\begin{center}
\caption[]{ Sensitivity of color $u-g$ to metallicity.}\label{Tab2}
 \begin{tabular}{ccccc}
  \hline\noalign{\smallskip}
Sensitivity    & Dwarf star    &    & Giant star    & \\
& $\mu_{1}$ (mag $\rm dex^{-1}$)    & $\sigma_{1}$ (mag $\rm dex^{-1}$)    & $\mu_{2}$ (mag $\rm dex^{-1}$)    & $\sigma_{2}$ (mag $\rm dex^{-1}$) \\
  \hline\noalign{\smallskip}
All stars    & 0.1437    & 0.1373    & 0.1815    & 0.1539 \\ 
0.0 --- $+$0.5    & 0.3356    & 0.1388    & 0.4239    & 0.1849 \\
$-$0.5 ---    0.0    & 0.2798    & 0.1275    & 0.3534    & 0.1808 \\
$-$1.0 --- $-$0.5    & 0.2249    & 0.1288    & 0.2841    & 0.1593 \\
$-$1.5 --- $-$1.0    & 0.1431    & 0.0693    & 0.1808    & 0.0887 \\
$-$2.0 --- $-$1.5    & 0.1022    & 0.0647    & 0.1292    & 0.0788 \\
$-$3.0 --- $-$2.0    & 0.0442    & 0.0222    & 0.0559    & 0.0282 \\
$-$4.0 --- $-$3.0    & 0.0195    & 0.0148    & 0.0247    & 0.0176 \\
  \noalign{\smallskip}\hline
\end{tabular}
\end{center}
\tablecomments{1\textwidth}{Here $\mu_{i}$, $\sigma_{i}$ (i=1, 2) is the mean value and scatter of the sensitivity. The first row is the total sensitivity, the remaining rows are for different bins of metallicity with range specified.}
\end{table}

\section{Classification of the stars}

As illustrated in Section 3, color $NUV-u$ is thought as an indicator of surface gravity. In this section, we test the classification of stars by using color $NUV-u$. The original parameter space coverage is too sparse (0.5 dex for log $g$ \,$\ge 0$). We thus integrate the theoretical spectra to a step of 0.05\,dex by using the PHOENIX high-resolution library. 

We mainly focus on classification of FGK-type stars (4,500 to 6,600 K) in this work. To simulate a realistic situation, the observational surface gravity distribution (equal to luminosity function) is adopted from stellar samples of LAMOST\,DR8 (\url{http://www.lamost.org/dr8/v2.0/catalogue}). The distribution of log\,$g$ shown in Fig.\,6 is used to re-sample the number of used theoretical spectra and their colors. Furthermore, we enlarge our sample by repeated sampling for 5 times.

There are 13,916 stars in our final sample (2,177 giant stars with log $g$ \textless \,3.5 and 11,739 dwarf stars with log $g$ \,$\ge 3.5$), as shown in Fig.\,7. In order to select giant stars with a minimum contamination from dwarf stars, we empirically define the color cuts that roughly follow the iso-number-ratio of 60\% in the contour. This yields
\begin{equation}\label{eq4}
(NUV-u) \ge 0.375 (g-i)+0.5212
\end{equation}

\begin{equation}\label{eq5}
(NUV-u) \ge 0.462 (g-i)+0.469
\end{equation}

With the above cuts, 1,069 stars are selected as giant stars (with 815 dwarf stars as contamination). The completeness of the selected giant stars is therefore around 49\% (1,069/2,177), along with a 57\% (1,069/1,884) purity.

 \begin{figure}
   \centering
  \includegraphics[width=14cm, angle=0]{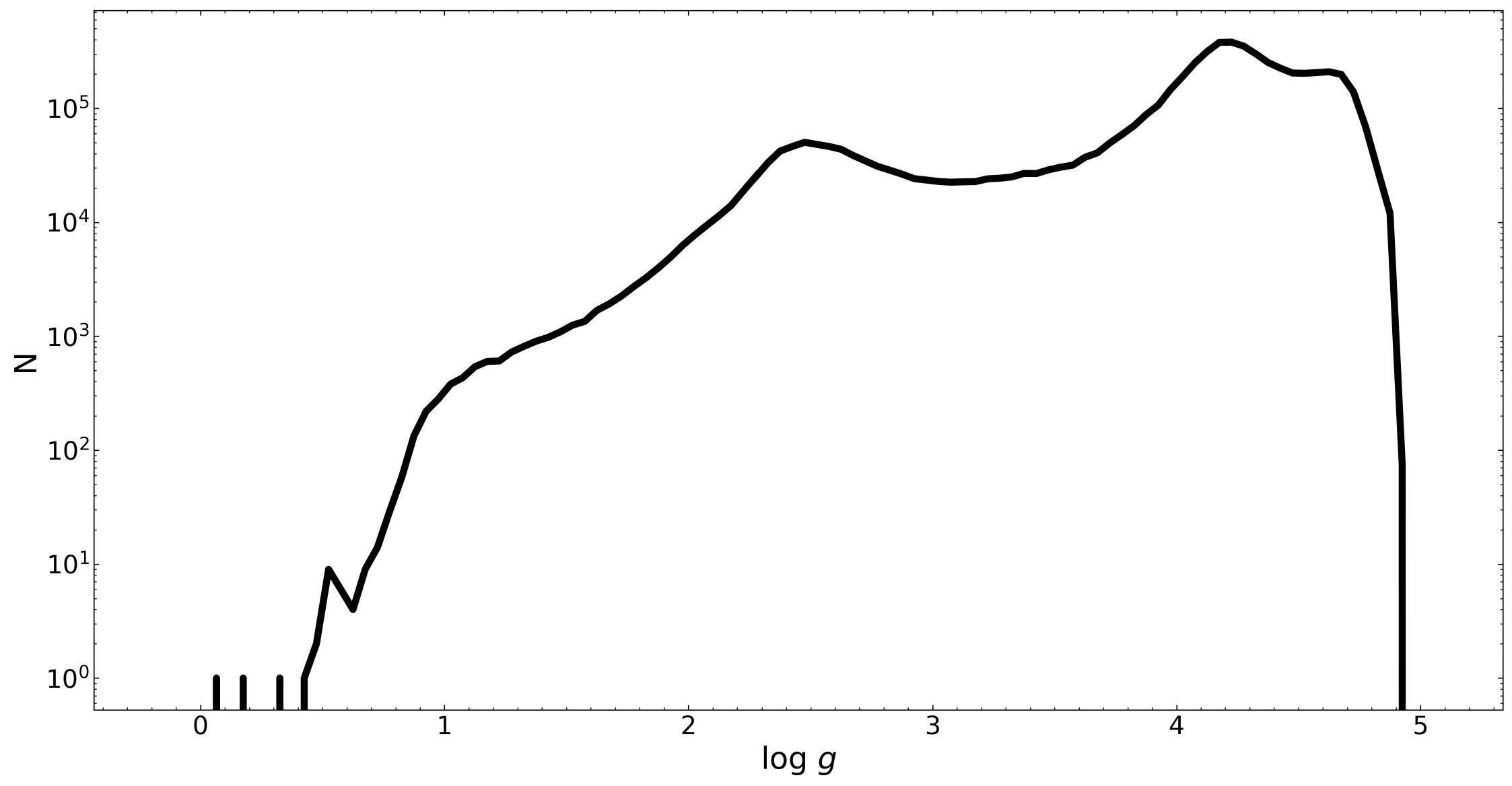}
   \caption{The distribution of stellar surface gravity log $g$ revealed from LAMOST DR8.}
   \label{Fig6}
   \end{figure}

\begin{figure}
   \centering
  \includegraphics[width=14cm, angle=0]{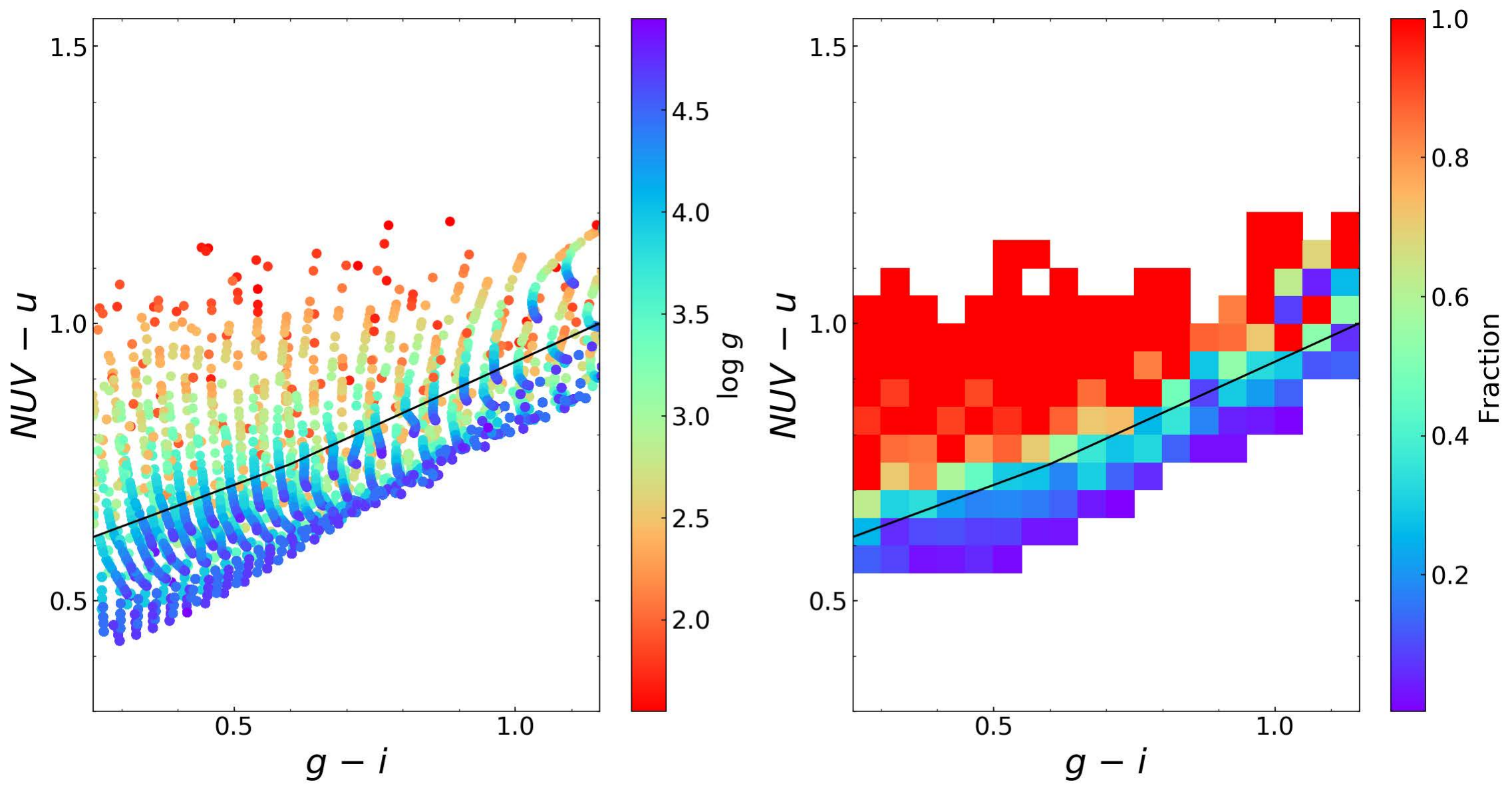}
   \caption{Left panel: ($NUV$-$u$) versus ($g$-$i$) of the stars, color coded by log $g$, as shown by the right color bars. Right panel: distribution of giant stars (log $g$ \textless \,3.5) in the ($NUV$-$u$) versus ($g$-$i$) plane (with a bin size of 0.05 mag in each axis). Black lines in both panels denote the cuts that we apply to distinguish dwarf and giant stars.}
   \label{Fig7}
   \end{figure}

Dwarf stars are chosen with the opposite color cuts that yields
\begin{equation}\label{eq6}
(NUV-u) < 0.375 (g-i)+0.5212
\end{equation}

\begin{equation}\label{eq7}
(NUV-u) < 0.462 (g-i)+0.469
\end{equation}

With the above cuts, 10,924 stars are selected as dwarf stars (with 1,108 giant stars as contamination). That means the completeness of the selected dwarf stars is around 93\% (10,924/11,739), along with a 91\% (10,924/12,032) purity. We note that similar efforts have been explored using colors from the SDSS and SkyMapper surveys (e.g., \citealt{Huang+etal+2019, Zhang+etal+2021}).

\section{Tests on performances of estimating metallicity}

\begin{figure}
   \centering
  \includegraphics[width=14cm, angle=0]{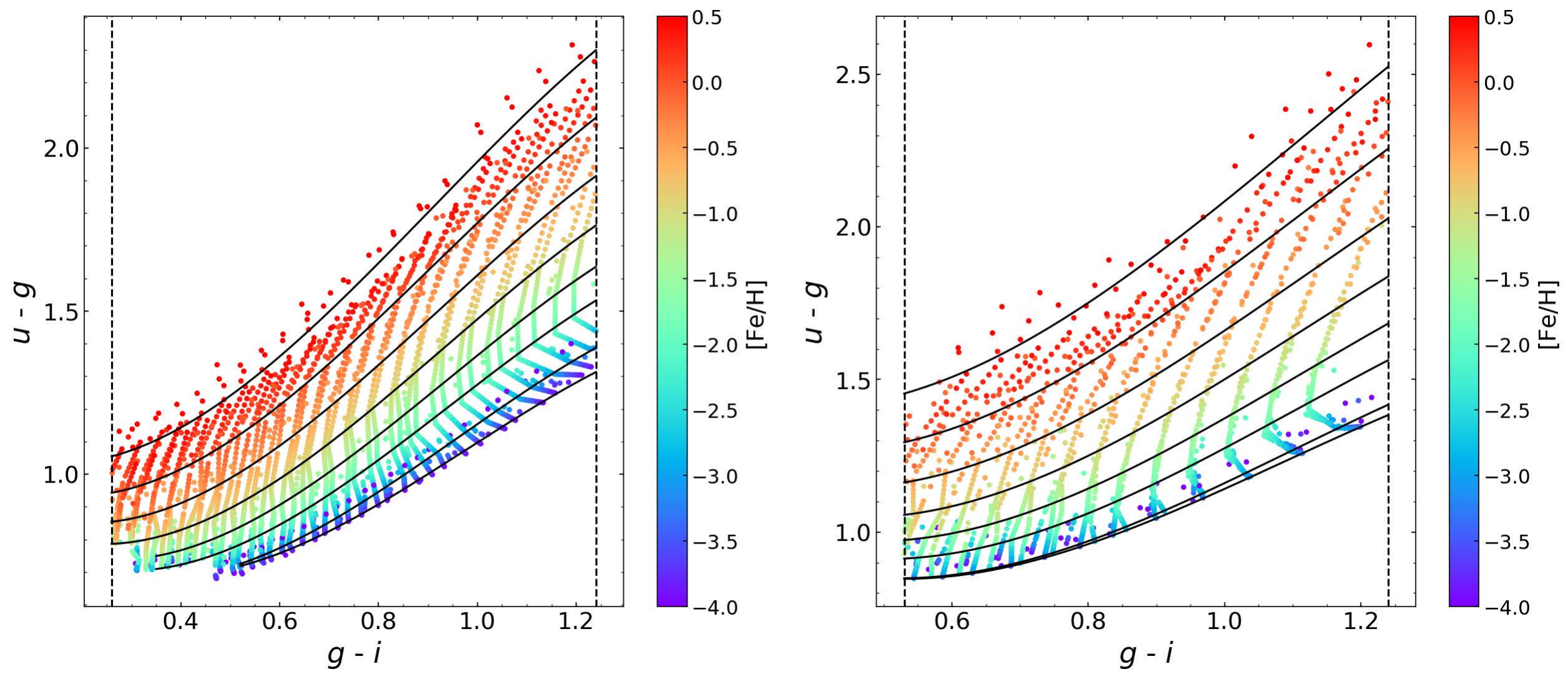}
   \caption{Distribution of the dwarf stars (left panel) and giant stars (right panel) in the ($u$-$g$) versus ($g$-$i$) plane, color coded by metallicities ([Fe/H]), as shown by the right color bars. The black lines represent our best fits for different values of [Fe/H], as described by Equation 4. From top to bottom, the values of [Fe/H] are $+$0.5, 0.0, $-$0.5, $-$1.0, $-$1.5, $-$2.0, $-$3.0 and $-$4.0, respectively. The dashed lines mark the color region that the fits are adopted.}
   \label{Fig8}
   \end{figure}

The main purpose of this section is to figure out how accurate metallicity can be measured from the CSST broad-band colors. The original parameter space coverage is too sparse (0.5 dex for [Fe/H] \,$\ge -2.0$, 1 dex for [Fe/H]\,$< -2.0$). We thus integrate the theoretical spectra to a step of 0.05\,dex. We then try to construct the metallicity-dependent stellar loci of $u-g$ versus $g-i$; this is the basic relation for further metallicity estimates. To achieve a realistic stellar loci, the observational surface gravity distribution shown in Fig.\,6 is again used to re-sample the number of used theoretical spectra and their colors. We classify them into two categories: dwarf stars with log $g \ge$ 3.5 and giant stars with log $g <$ 3.5.

The first and the most important step is to construct the metallicity-dependent stellar loci. As shown in Fig.\,8., the sequences of different metallicities ranging from [Fe/H] = $-$4 to [Fe/H] = $+$0.5, as color ($u-g$) changing with ($g-i$), can be seen clearly for both dwarf and giant stars. Similar to previous studies (\citealt{Yuan+etal+2015, Huang+etal+2022, Huang+etal+2023, Lin+etal+2022}), third-order 2-D polynomials are adopted to fit the color ($u-g$), as a function of ($g-i$) and [Fe/H] for dwarf and giant stars, respectively:
\begin{equation}\label{eq8}
  {(u-g)} = a_{0,0}+a_{0,1}y+a_{0,2}y^{2}+a_{0,3}y^{3}+a_{1,0}x+a_{1,1}xy+a_{1,2}xy^{2}+a_{2,0}x^{2}+a_{2,1}x^{2}y+a_{3,0}x^{3},
\end{equation}
where $x$ denotes ($g$-$i$) and $y$ denotes [Fe/H]. Three-sigma clipping is applied during the fitting process and the fit coefficients are listed in Table 3. Using the empirical stellar loci defined above, the maximum-likelihood approach are then adopted to derive metallicity estimates. For a given star, the likelihood is expressed as:
\begin{equation}\label{eq9}
  {L_{c}} = \frac{1}{\sqrt{2\pi}\sigma_{\rm c_{PHOE}}}exp{\frac{-(\rm c_{PHOE}-\rm c_{pred})^{2}}{2\sigma_{\rm c_{PHOE}}^{2}}}.
\end{equation}
where $\rm c_{PHOE}$ = ($u-g$) is assumed to be independent Gaussian variables calculated using PHOENIX spectra, $\rm c_{pred}$ is the same color that can be predicted from our metallicity-dependent stellar loci (i.e., Equation 8). The value of [Fe/H] is varied from $-$4.0 to $+$0.5, with steps of 0.05 dex, when predicting ($u-g$). With the likelihood function, the best fit color ($u-g$) can be derived for each star, the value of [Fe/H] corresponding to this best fit color is taken as the predicted [Fe/H]. It is worth noting that applicability of ($g-i$) range exists in the current method of estimating stellar metallicity. It is 0.26 to 1.24, 0.53 to 1.24 for dwarf and giant stars, respectively. Moreover, the applicable upper and lower limits of [Fe/H] in color ($u-g$) are defined for various ($g-i$) bins.
   
CSST, as a planned large-scale deep survey, owns powerful capability of measuring stellar atmospheric parameters, as supported by our mock data tests. It can be seen that predicted metallicity from our metallicity-dependent stellar loci is in good agreement with metallicity from PHOENIX spectra for [Fe/H] \textgreater $-$2 for both dwarf and giant stars, with typical values around 0.17 dex and 0.20 dex, respectively. For dwarf stars (as shown in Fig.\,9), the precision of the predicted metallicity is 0.14 dex if colors $u-g$, $g-i$ with a 0.001 mag random error. Even if the color error is up to 0.015 mag, the scatter is still smaller than 0.21 dex, which can be comparable to that of medium-resolution spectroscopy. Although the scatter becomes larger with increasing color error, the metal-rich parts ([Fe/H] \textgreater $-$2 ) still shows great agreement with the true value. Fig.\,10 shows the results on the performances of estimating metallicity for giant stars. The scatter is smaller than 0.23 dex when the random error of colors lies within 0.015 mag, though a bit larger than that from dwarf stars under the same color error. Good agreement also exists even under a color error of 0.025 mag.

Correspondingly, using the tool ``Exposure time calculator for space telescope''\footnote{https://nadc.china-vo.org/csst-bp/etc-ms/etc.jsp}, the $g$ band limiting magnitudes under various color errors are derived, as listed in Table 4. We expect the metallicity estimate is better than 0.20 dex for stars with $g$-band magnitude down to 20.8 (color error smaller than 0.01 mag).

\begin{figure}
   \centering
  \includegraphics[width=14cm, angle=0]{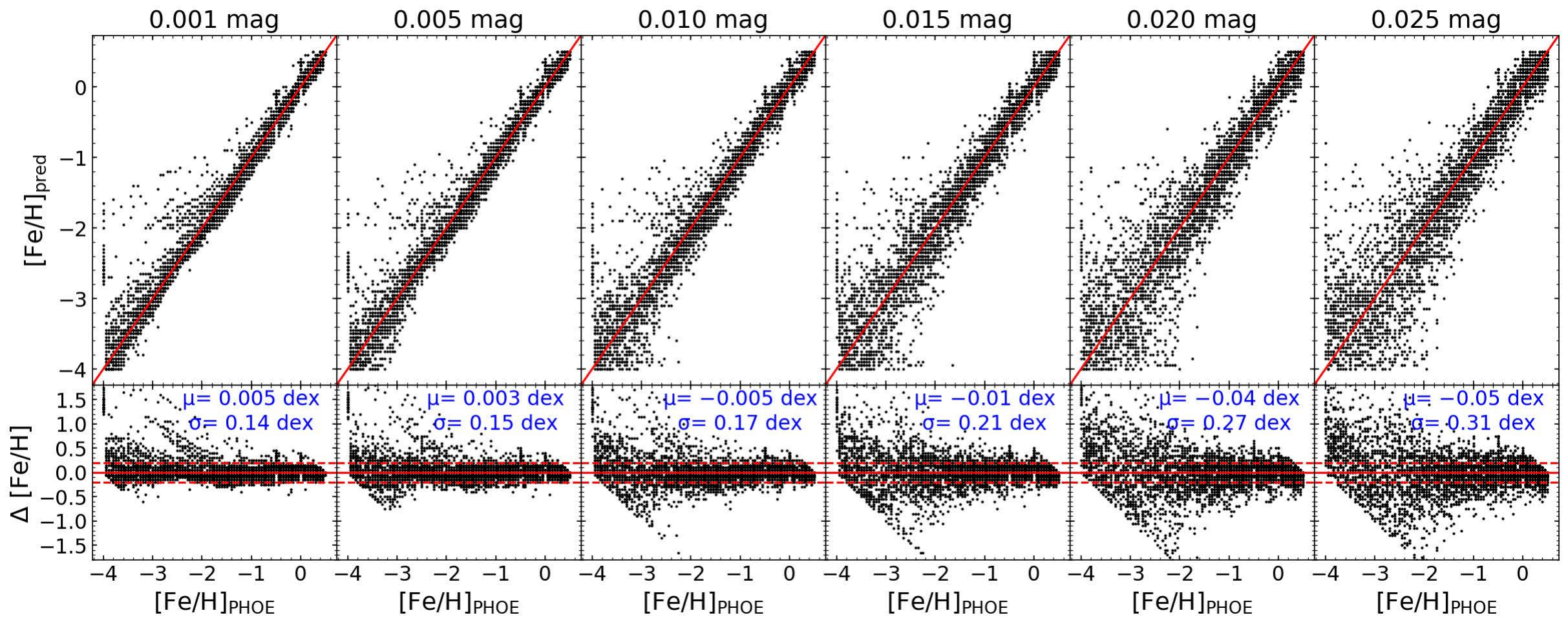}
   \caption{Comparisons of predicted metallicity and metallicity from PHOENIX spectra of dwarf stars for CSST filter systems. The lower pars of each panel is the metallicity difference (predicted metallicity minus metallicity from PHOENIX spectra) as a function of metallicity from PHOENIX spectra. $\Delta$ [Fe/H] = $\pm{0.2}$ dex lines are plotted to guide the eye. Panels are arranged as a function of random error of color (from the left to the right panels).}
   \label{Fig9}
   \end{figure}

\begin{figure}
   \centering
  \includegraphics[width=14cm, angle=0]{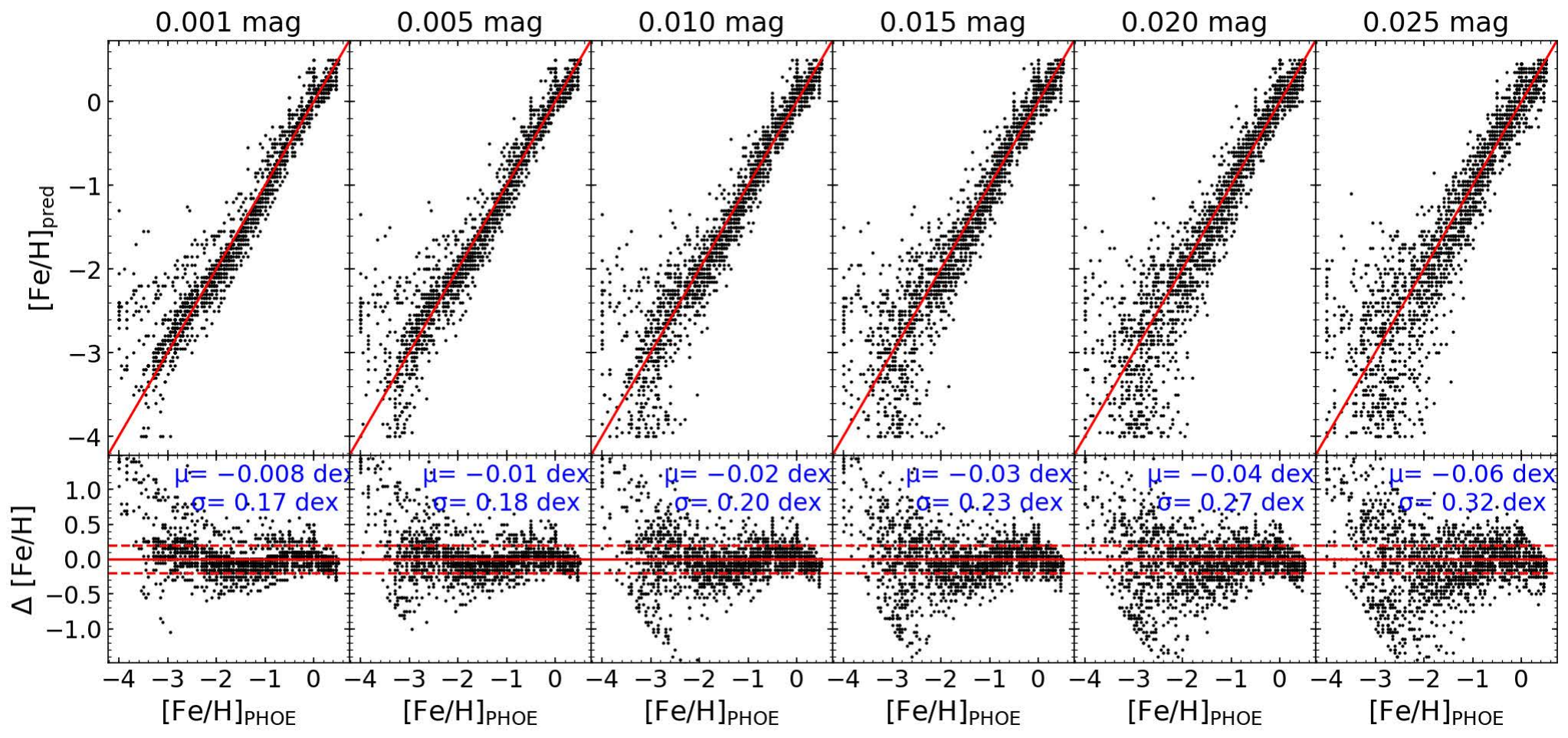}
   \caption{Similar to Fig.9 but for giant stars.}
   \label{Fig10}
   \end{figure}

\begin{table}
\begin{center}
\caption[]{ Coefficients for metallicity and effective temperature estimates for dwarf and giant Stars.}\label{Tab3}
\begin{tabular}{clccc}
  \hline\noalign{\smallskip}
Coeff.    & ($u-g\rm)_{dwarf}$    & ($u-g\rm)_{giant}$    & $T\rm_{eff,dwarf}$    & $T\rm_{eff,giant}$ \\
  \hline\noalign{\smallskip}
$a_{0,0}$    & 1.00914783    & 1.75846485    & 0.66217487    & 0.69958833 \\
$a_{0,1}$    & 0.13730156    & 0.16147817    & $-$0.00643389    & $-$0.01426616 \\
$a_{0,2}$    & 0.04259072    & 0.03699292    & $-$0.01095072    & $-$0.01687791 \\ 
$a_{0,3}$    & 0.00248219    & 0.00270348    & --    & -- \\
$a_{1,0}$    & $-$0.84918537    & $-$2.56274845    & 0.35698828    & 0.28526773 \\
$a_{1,1}$    & 0.25060675    & 0.22458836    & $-$0.03341913    & $-$0.04121234 \\
$a_{1,2}$    & 0.01201316    & 0.03500375    & --    & -- \\
$a_{2,0}$    & 2.52952155    & 3.77376427    & $-$0.0046996    & 0.02679721 \\
$a_{2,1}$    & $-$0.04099565    & 0.03681587    & --    & -- \\
$a_{3,0}$    & $-$0.91951624    & $-$1.11547021    & --    & -- \\
  \noalign{\smallskip}\hline
\end{tabular}
\end{center}
\end{table}

\begin{table}
\begin{center}
\caption[]{ Limiting magnitude of $g$ band under different color errors.}\label{Tab4}
 \begin{tabular}{ccccccc}
  \hline\noalign{\smallskip}
Color error    & 0.001    & 0.005    & 0.01    & 0.015    & 0.02    & 0.025 \\
  \hline\noalign{\smallskip} \\
Limiting magnitude    & 15.85    & 19.31    & 20.77    & 21.59    & 22.13    & 22.57 \\
  \noalign{\smallskip}\hline
\end{tabular}
\end{center}
\end{table}

\begin{table}
\begin{center}
\caption[]{ Coefficients for $T\rm_{eff}$ estimates for dwarf and giant Stars from $g-z$ and $g-y$ colors.}\label{Tab5}
\begin{tabular}{ccccc}
  \hline\noalign{\smallskip}
Coeff.    & $T\rm_{dwarf,g-z}$    & $T\rm_{giant,g-z}$    & $T\rm_{dwarf,g-y}$    & $T\rm_{giant,g-y}$ \\
  \hline\noalign{\smallskip}
$a_{0,0}$    & 0.69098643    & 0.71306772    & 0.69702855     & 0.70894565 \\
$a_{0,1}$    & $-$0.00190293    & $-$0.01194074    & $-$0.00042429    & $-$0.0150918 \\
$a_{0,2}$    & $-$0.00897226    & $-$0.01468411    & $-$0.00924179    & $-$0.01249024 \\ 
$a_{1,0}$    & 0.29964223    & 0.26189104    & 0.299407    & 0.26657628 \\
$a_{1,1}$    & $-$0.02588283    & $-$0.02968957    & $-$0.02339048    & $-$0.01839136 \\
$a_{2,0}$    & $-$0.00345118    & 0.00980925    & $-$0.01344877    & 0.00278537 \\
  \noalign{\smallskip}\hline
\end{tabular}
\end{center}
\end{table}

\section{Tests on performances of estimating effective temperature}

\begin{figure}
   \centering
  \includegraphics[width=14cm, angle=0]{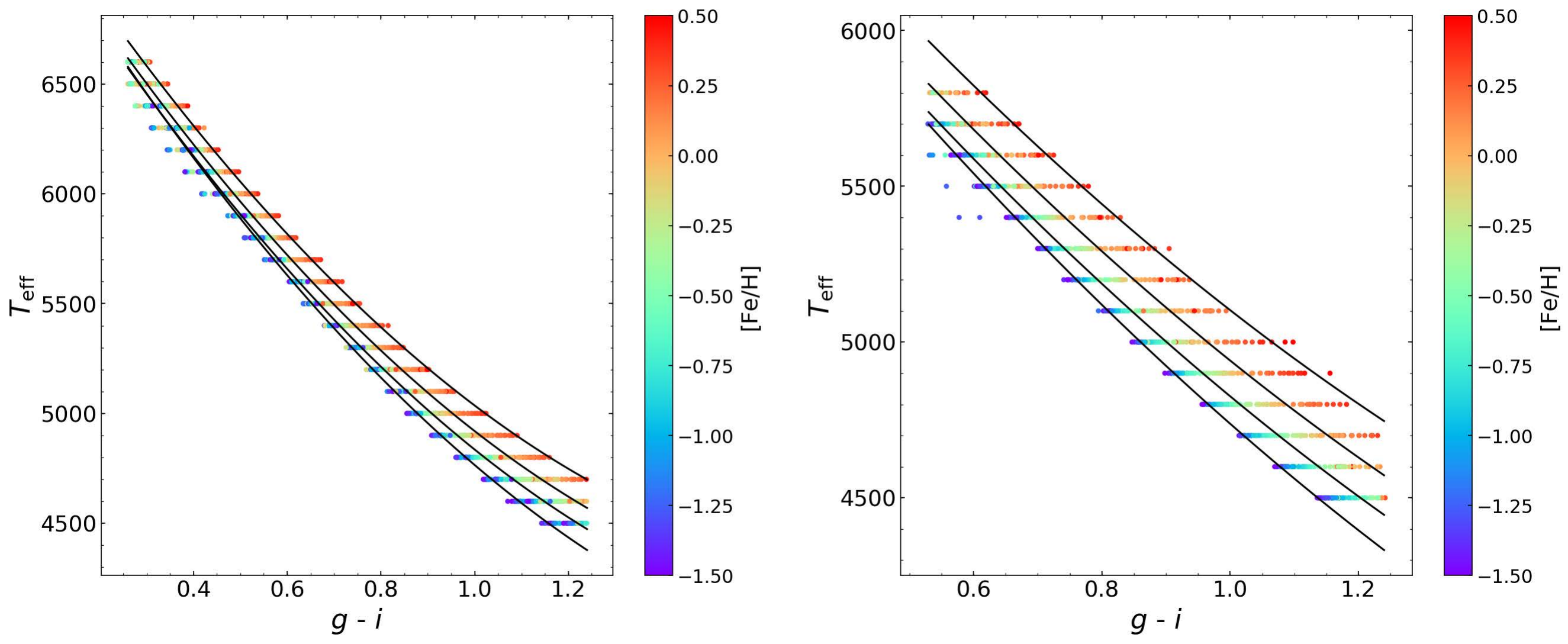}
   \caption{Effective temperature, as a function of color ($u$-$g$), for the dwarf (left panel) and giant (right panel) stars color coded by metallicity, as shown in the right color bars. The black lines represent our best fits for different values of [Fe/H], as described by Equation 6. From top to bottom, the values of [Fe/H] are $+$0.5, 0.0, $-$0.5 and $-$1.5, respectively.}
   \label{Fig11}
   \end{figure}

\begin{figure}
   \centering
  \includegraphics[width=14cm, angle=0]{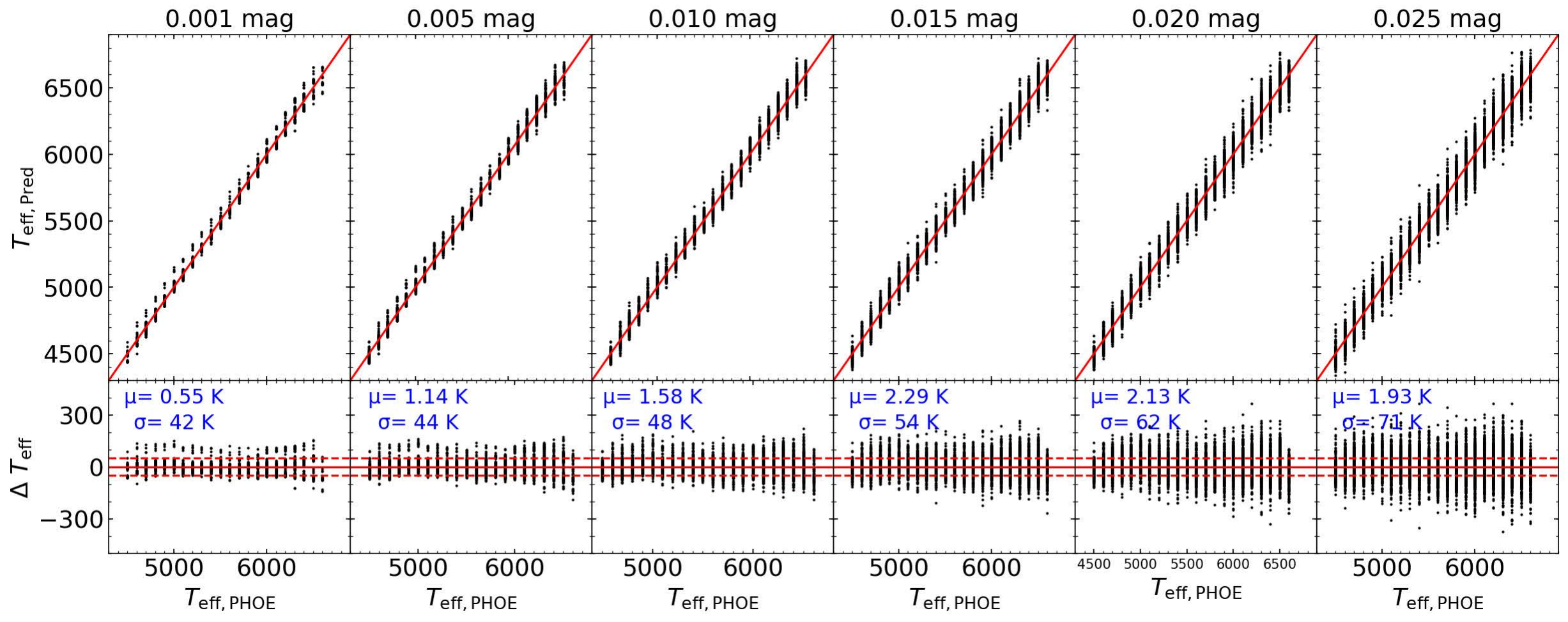}
   \caption{Comparisons of predicted effective temperature from $g-i$ color and effective temperature from PHOENIX spectra of dwarf stars for CSST filter systems. $\Delta$ $T\rm_{eff}$ = $\pm{50}$ K lines are plotted to guide the eye.}
   \label{Fig12}
   \end{figure}

\begin{figure}
   \centering
  \includegraphics[width=14cm, angle=0]{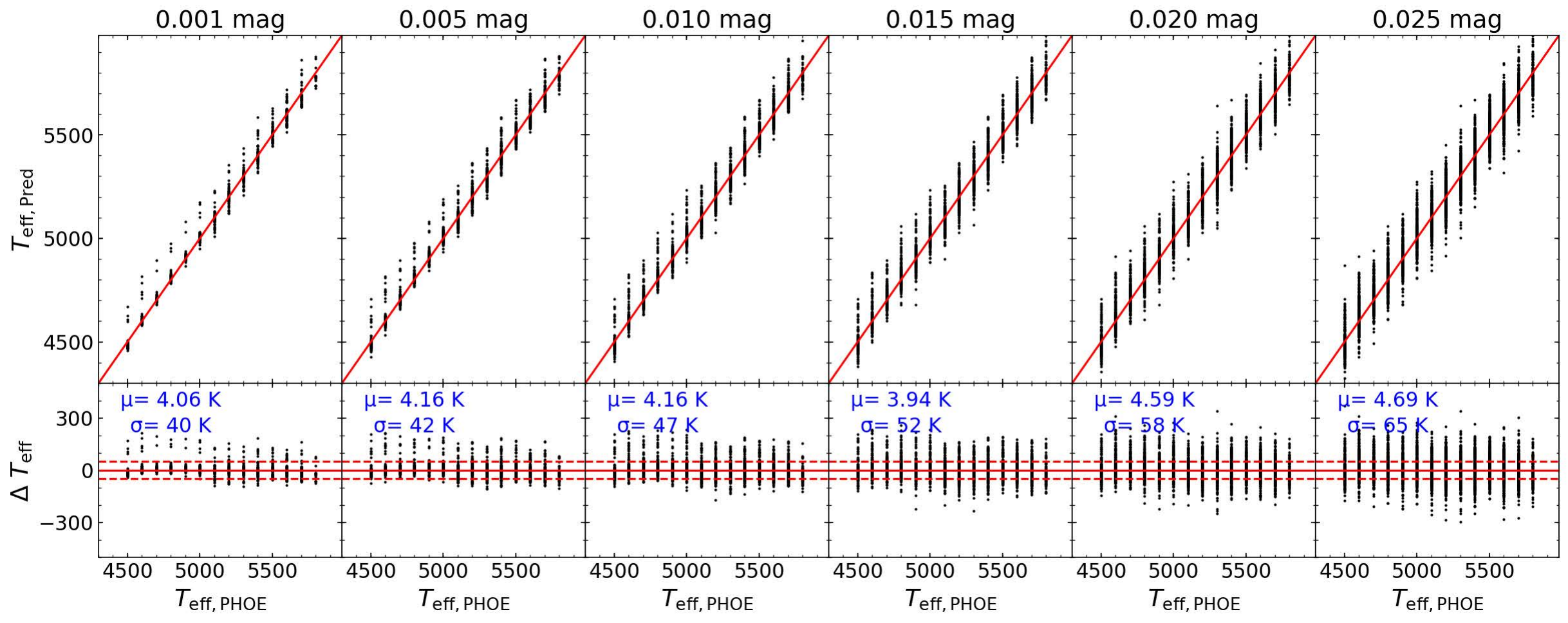}
   \caption{Similar to Fig.12 but for giant stars.}
   \label{Fig13}
   \end{figure}

\begin{figure}
   \centering
  \includegraphics[width=14cm, angle=0]{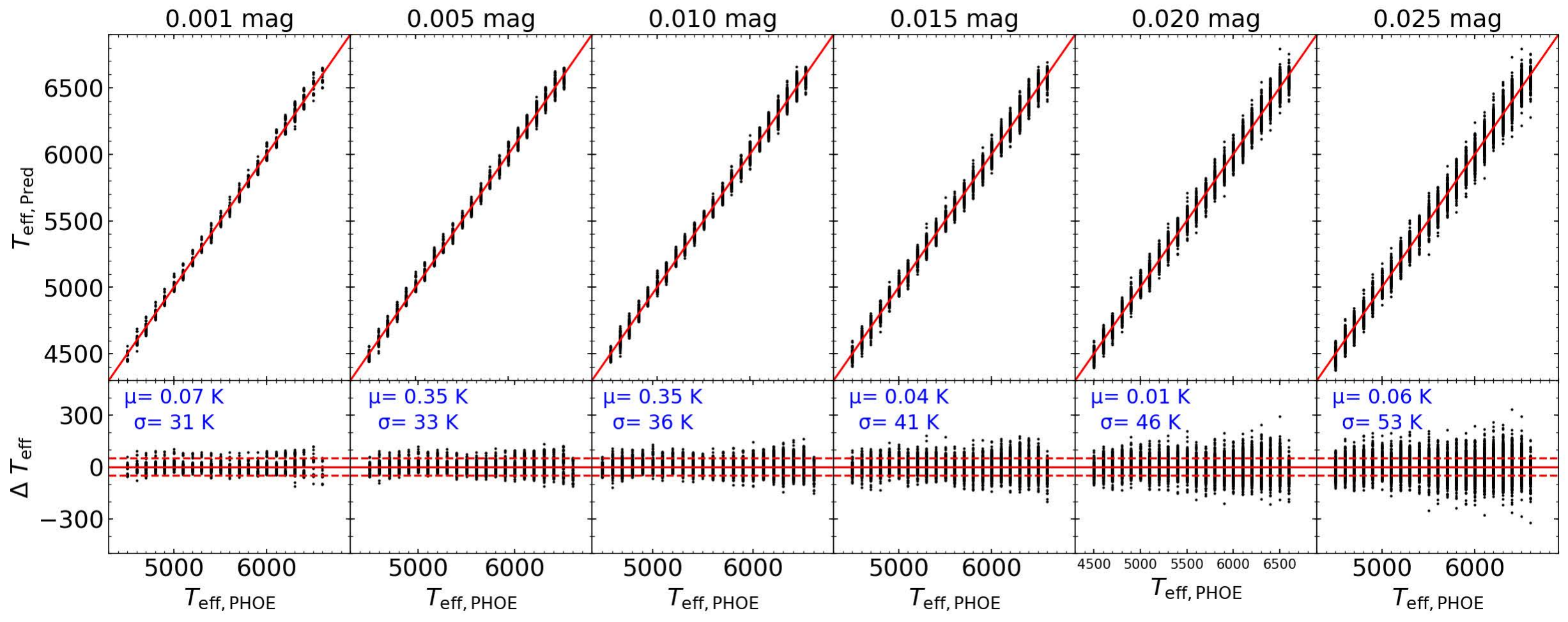}
   \caption{Similar to Fig.12 but for estimating effective temperature of dwarf stars from $g-z$ color.}
   \label{Fig14}
   \end{figure}

\begin{figure}
   \centering
  \includegraphics[width=14cm, angle=0]{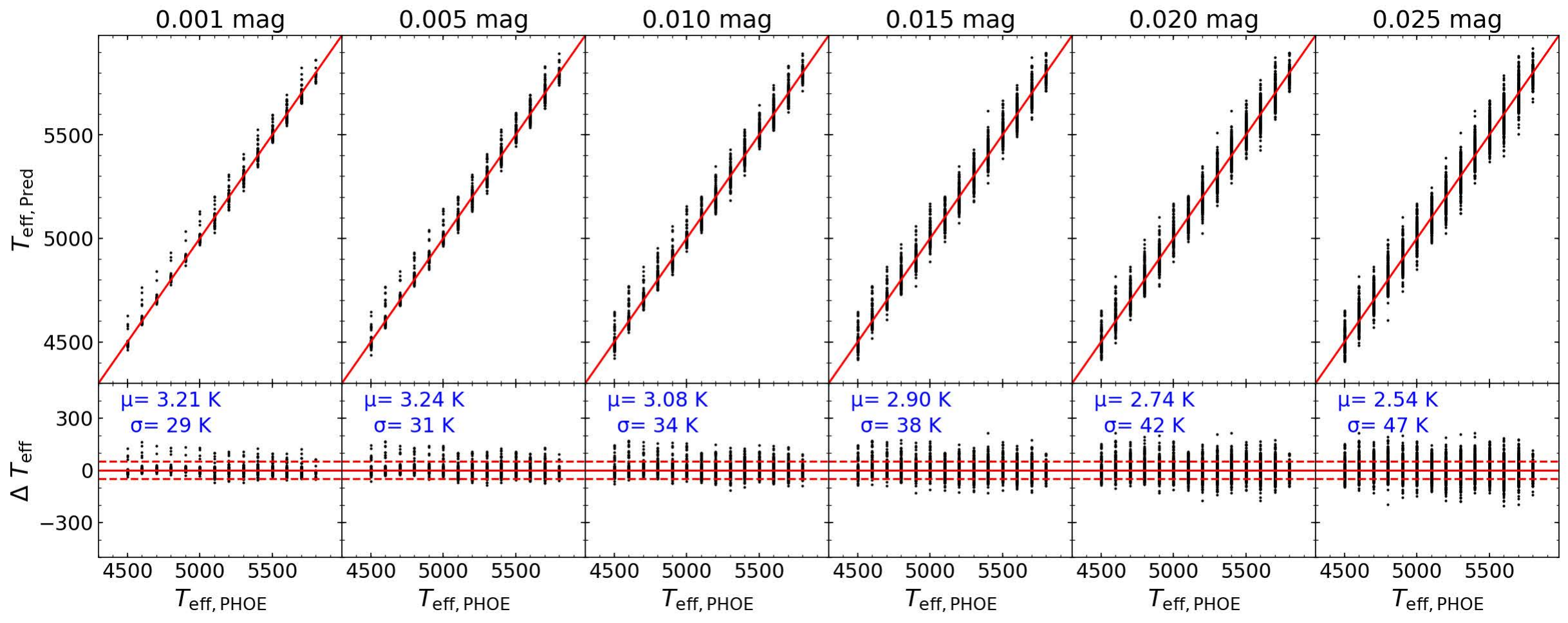}
   \caption{Similar to Fig.12 but for estimating effective temperature of giant stars from $g-z$ color}
   \label{Fig15}
   \end{figure}

\begin{figure}
   \centering
  \includegraphics[width=14cm, angle=0]{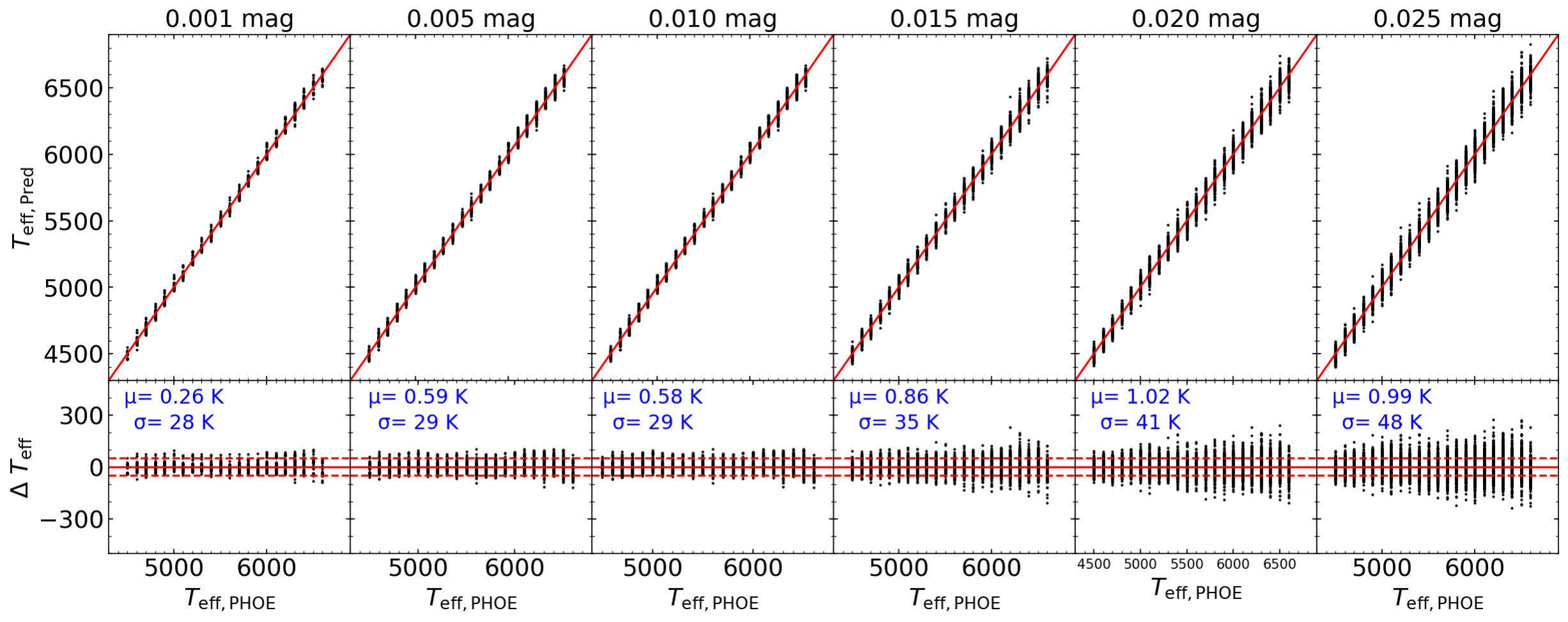}
   \caption{Similar to Fig.12 but for estimating effective temperature of dwarf stars from $g-y$ color.}
   \label{Fig16}
   \end{figure}

\begin{figure}
   \centering
  \includegraphics[width=14cm, angle=0]{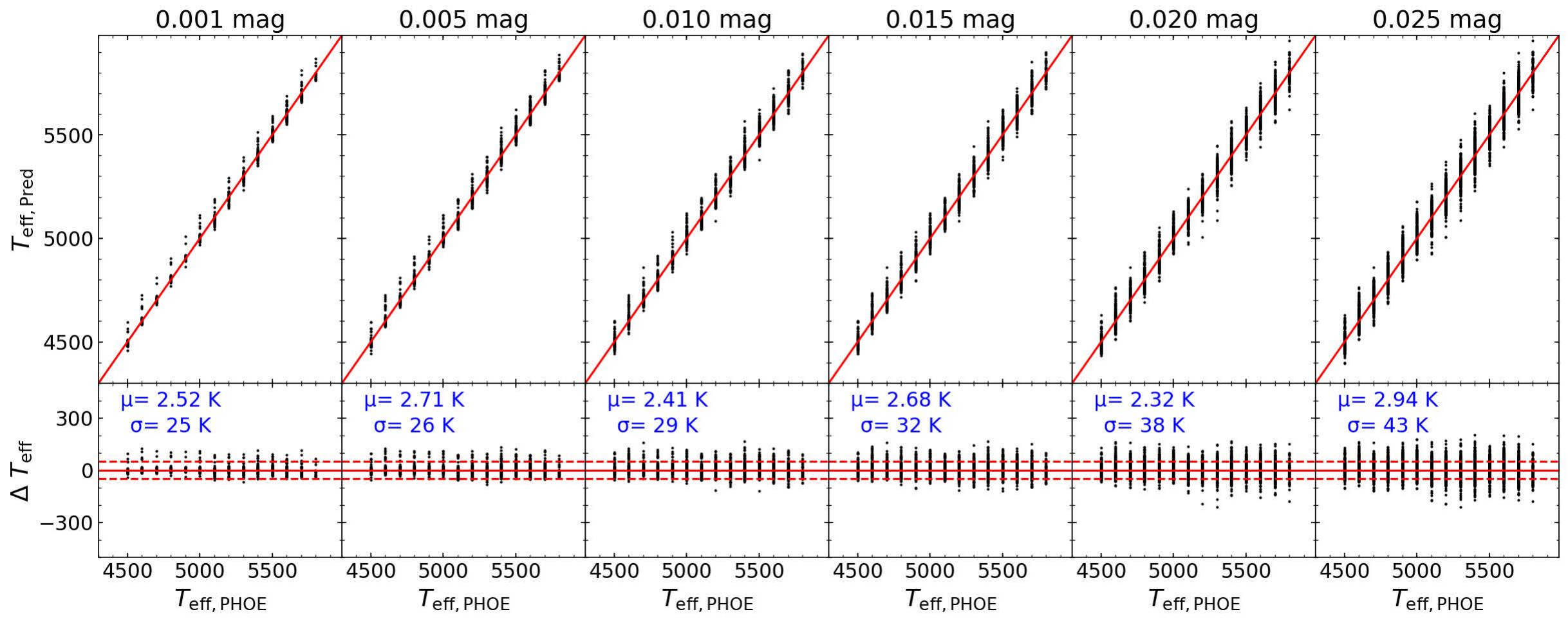}
   \caption{Similar to Fig.12 but for estimating effective temperature of giant stars from $g-y$ color}
   \label{Fig17}
   \end{figure}

In this section, tests on performances of estimating effective temperature are done. Similar to that illustrated in Section 5, the essential one is to construct the metallicity-dependent $T\rm_{eff}$-color relations. To accomplish that, second-order 2-D polynomials are adopted to fit the data points for dwarf and giant stars, respectively (Fig.\,11):
\begin{equation}\label{eq10}
   \theta\rm_{eff}= a_{0,0}+a_{0,1}y+a_{0,2}y^{2}+a_{1,0}x+a_{1,1}xy+a_{2,0}x^{2}.
\end{equation}
where $\theta\rm_{eff}$ = 5,000/$T\rm_{eff}$, $x$ denotes ($g$-$i$) and $y$ denotes [Fe/H]. Three-sigma clipping is applied in the fitting process. The fit coefficients are listed in Table 3. 

The results on performances of estimating effective temperature from $g-i$ color are shown in Figs.\,12 and 13. High precision of effective temperature is achieved for CSST broad-band photometry, with typical values around 48 K and 47 K for dwarf and giant stars, respectively. For tests from dwarf stars,the precision of the predicted effective temperature lies within 50 K when the random error of color is smaller than 0.01 mag. Even for the maximum color error in our mock data tests, the precision is high as well, with the value around 70 K. For giant stars, the precision could be a bit better, which is smaller than 47 K even if a random color error of 0.01 mag is given. This again shows the powerful capability of measuring stellar atmospheric parameters from CSST survey.

Moreover, tests on estimates of effective temperature from $g-z$ and $g-y$ colors are also done using the similar metallicity-dependent $T\rm_{eff}$-color relations, as described by Equation 10. Figs.\,14 and 15 show the comparisons between $T\rm_{eff}$ predicted from $g-z$ color and the true $T\rm_{eff}$ for the dwarf and giant stars, respectively. Good agreement can be clearly seen. Compared to $T\rm_{eff}$ predicted from $g-i$ color, $T\rm_{eff}$ predicted from $g-z$ color achieves a higher precision. 

Figs.\,16 and 17 show the comparisons between $T\rm_{eff}$ predicted from $g-z$ color and the true $T\rm_{eff}$ for the dwarf and giant stars, respectively. The precision of predicted $T\rm_{eff}$ is further improved, with typical values around 29 K and 29 K for dwarf and giant stars, respectively. Therefore, in the future $T\rm_{eff}$ estimates from CSST survey, we recommend $g-y$ color to be the priority.

\begin{figure}
   \centering
  \includegraphics[width=14cm, angle=0]{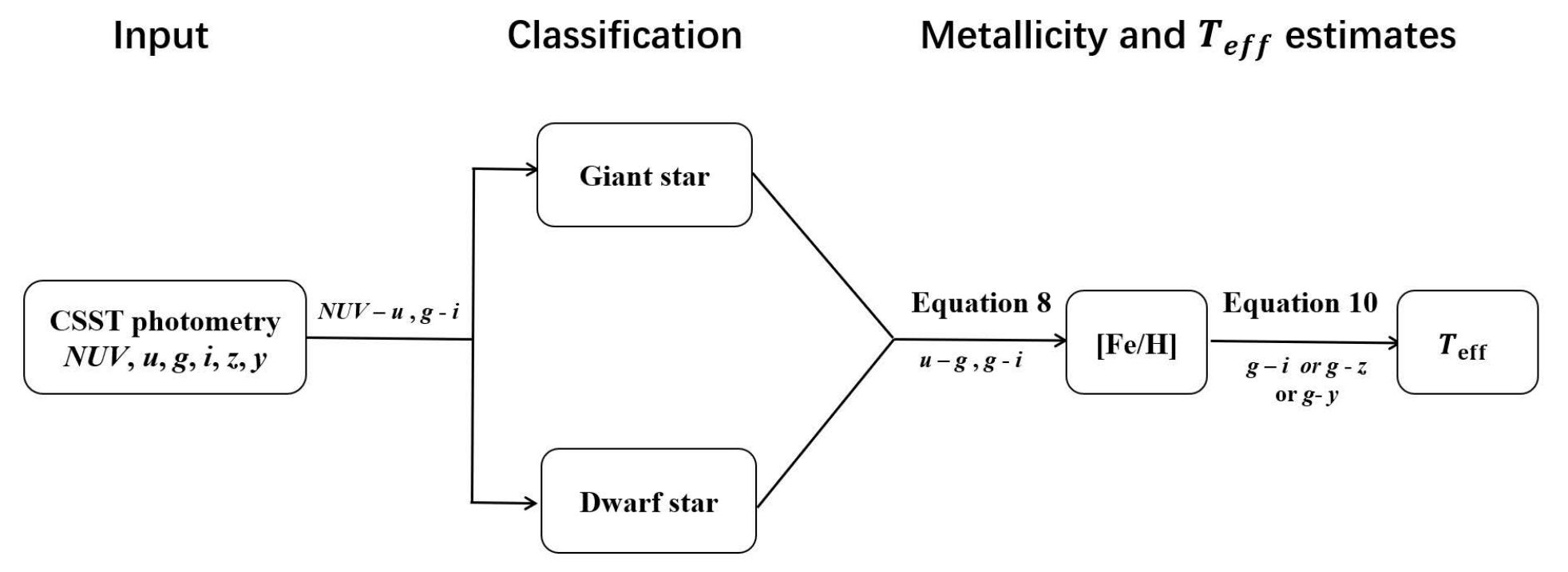}
   \caption{Flowchart of the code. Three modules are contained in our code including the input module, classification module, and the metallicity and efficient temperature estimates module.}
   \label{Fig18}
   \end{figure}

It is worth noting that uncertainties from photometric calibration and reddening correction are not taken into account in above analysis. Moreover, as discussed in Section 3, we may not make full use of the power of CSST broad-bands, especially the $NUV$ filter due to the imperfect performance of the PHOENIX theoretical spectra. In the near future, we will continue to improve these shortcomings to present more realistic analysis of CSST broad-bands on deriving stellar parameters.

Finally, we release a code\footnote{https://pypi.org/project/CSST-parameter/} to estimate stellar atmospheric parameters ([Fe/H], $T_{\rm eff}$) for CSST. Fig.\,18 shows the flowchart of our code.

\section{Summary}
Sensitivity of CSST broad-band colors to stellar atmospheric parameters (metallicity, [Fe/H], surface gravity, log $g$) is evaluated using synthetic colors integrated from model spectra. Result shows that colors from the optical and near-ultraviolet filter systems adopted by CSST survey show significant sensitivities to the stellar atmospheric parameters, in particular the metallicity. Sensitivity of color to metallicity of giant stars is higher than that from dwarf stars, although with a little bit larger scatter.

According to our mock data tests, dwarf stars can be distinguished from giant stars. CSST can offer accurate stellar metallicity and effective temperature estimates for stars with $g$-band magnitude down to 20.8, with the high precision around 0.20 dex and 50 K, respectively. It is inspiring that the assembly history of the Milky Way can be further revealed using such a large sample of stars with accurate estimates of stellar atmospheric parameters.

\begin{acknowledgements}
We acknowledge the science research grants from the China Manned Space Project with NO. CMS-CSST-2021-A08. Y.H. acknowledges the NSFC for grant Nos. 11903027 and 11833006 and the National Key RD Program of China for
grant No. 2019YFA0405503. H.W.Z. acknowledges the science research grants from the China Manned Space Project, NSFC
for grant Nos. 11973001, 12090040, and 12090044, and the
National Key RD Program of China for grant No. 2019YFA0405504.
\end{acknowledgements}


\begin{thebibliography}{}

  \bibitem[Beers \& Christlieb(2005)]{Beers+Christlieb+2005} Beers, T. C. \& Christlieb, N. 2005, \araa, 43, 531

  \bibitem[Benitez et al.(2014)]{Benitez+etal+2014} Benitez, N., Dupke, R., Moles, M., et al. 2014, arXiv:1403.5237

  \bibitem[Bonifacio et al.(2021)]{Bonifacio+etal+2021} Bonifacio, P., Monaco, L., Salvadori, S., et al. 2021, \aap, 651, A79

  \bibitem[Casagrande et al.(2010)]{Casagrande+etal+2010} Casagrande, L., Ram{\'\i}rez, I., Mel{\'e}ndez, J., et al. 2010, \aap, 512, A54

  \bibitem[Casagrande \& VandenBerg(2014)]{Casagrande+VandenBerg+2014} Casagrande, L. \& VandenBerg, D. A. 2014, \mnras, 444, 392

  \bibitem[Cenarro et al.(2019)]{Cenarro+etal+2019} Cenarro, A. J., Moles, M., Crist\'{o}bal-Hornillos, D., et al. 2019, \aap, 622, A176

  \bibitem[Chambers et al.(2016)]{Chambers+etal+2016} Chambers, K.~C., Magnier, E.~A., Metcalfe, N., et al. 2016, arXiv:1612.05560

  \bibitem[Fan et al.(2023)]{Fan+etal+2023} Fan, Z., Zhao, G., Wang, W., et al. 2023, arXiv:2306.15611

  \bibitem[Gaia Collaboration et al.(2016)]{Brown+etal+2016} Gaia Collaboration, Brown, A.~G.~A., Vallenari, A., et al. 2016, \aap, 595, A2

  \bibitem[Heger \& Woosley(2002)]{Heger+Woosley+2002} Heger, A. \& Woosley, S. E. 2002, \apj, 567, 532

  \bibitem[Heger \& Woosley(2010)]{Heger+Woosley+2010} Heger, A. \& Woosley, S. E. 2010, \apj, 724, 341

  \bibitem[Huang et al.(2015)]{Huang+etal+2015} Huang, Y., Liu, X.-W., Yuan, H.-B., et al. 2015, \mnras, 454, 2863

  \bibitem[Huang et al.(2019)]{Huang+etal+2019} Huang, Y., Chen, B.-Q., Yuan, H.-B., et al. 2019, \apjs, 243, 7

  \bibitem[Huang et al.(2022)]{Huang+etal+2022} Huang, Y., Beers, T. C., Wolf, C., et al. 2022, \apj, 925, 164

  \bibitem[Huang et al.(2023)]{Huang+etal+2023} Huang, Y., Beers, T.~C., Yuan, H.-B., et al. 2023, arXiv:2307.04469

  \bibitem[Husser et al.(2013)]{Husser+etal+2013} Husser, T.-O., Wende-von Berg, S., Dreizler, S., et al. 2013, \aap, 553, A6.

  \bibitem[Ishigaki et al.(2018)]{Ishigaki+etal+2018} Ishigaki, M. N., Tominaga, N., Kobayashi, C., \& Nomoto, K. 2018, \apj, 857, 46

  \bibitem[Ishigaki et al.(2021)]{Ishigaki+etal+2021} Ishigaki, M. N., Hartwig, T., Tarumi, Y., et al. 2021, \mnras, 506, 5410

  \bibitem[Ivezi\'{c} et al.(2008)]{Ivezic+etal+2008} Ivezi\'{c}, {\v{Z}}., Sesar, B., Juri\'{c}, M., et al. 2008, \apj, 684, 287

  \bibitem[Lardo et al.(2021)]{Lardo+etal+2021} Lardo, C., Mashonkina, L., Jablonka, P., et al. 2021, \mnras, 508, 3068

  \bibitem[Limongi \& Chieffi(2012)]{Limongi+Chieffi+2012} Limongi, M. \& Chieffi, A. 2012, \apjs, 199, 38

  \bibitem[Lin et al.(2022)]{Lin+etal+2022} Lin, J., Casagrande, L., \& Asplund, M. 2022, \mnras, 510, 433

  \bibitem[Lu et al.(2023)]{Lu+etal+2023} Lu, X., Yuan, H., Xu, S., et al. 2023, arXiv:2311.16901

  \bibitem[Mendes de Oliveira et al.(2019)]{Mendes de Oliverira+etal+2019} Mendes de Oliveira, C., Ribeiro, T., Schoenell, W., et al. 2019, \mnras, 489, 241

  \bibitem[Nomoto et al.(2013)]{Nomoto+etal+2013} Nomoto, K., Kobayashi, C., \& Tominaga, N.\ 2013, \araa, 51, 457

  \bibitem[Onken et al.(2019)]{Onken+etal+2019} Onken, C. A., Wolf, C., Bessell, M. S., et al. 2019, \pasa, 36, e033

  \bibitem[Ram{\'\i}rez \& Mel{\'e}ndez(2005)]{Ramirez+Melendez+2005} Ram{\'\i}rez, I. \& Mel{\'e}ndez, J. 2005, \apj, 626, 446

  \bibitem[Starkenburg et al.(2017)]{Starkenburg+etal+2017} Starkenburg, E., Martin, N., Youakim, K., et al. 2017, \mnras, 471, 2587

  \bibitem[Umeda \& Nomoto(2002)]{Umeda+Nomoto+2002} Umeda, H. \& Nomoto, K. 2002, \apj, 565, 385.

  \bibitem[Whitten et al.(2021)]{Whitten+etal+2021} Whitten, D. D., Placco, V. M., Beers, T. C., et al. 2021, \apj, 912, 147

  \bibitem[Wolf et al.(2018)]{Wolf+etal+2018} Wolf, C., Onken, C. A., Luvaul, L. C., et al. 2018, \pasa, 35, e010

  \bibitem[York et al.(2000)]{York+etal+2000} York, D. G., Adelman, J., Anderson, J. E., et al. 2000, \aj, 120, 1579

  \bibitem[Yuan et al.(2015)]{Yuan+etal+2015} Yuan, H., Liu, X., Xiang, M., et al. 2015, \apj, 799, 134

  \bibitem[Zepeda et al.(2022)]{Zepeda+etal+2022} Zepeda, J., Rasmussen, K. C., Beers, T. C., et al. 2022, \apj, 927, 13

  \bibitem[Zhan (2011)]{Zhan+2011} Zhan, H. 2011, SCIENTIA SINICA Physica, Mechanica \& Astronomica, 41, 1441 

  \bibitem[Zhang et al.(2021)]{Zhang+etal+2021} Zhang, R. Y., Yuan, H. B., Liu, X. W., et al. 2021, Research in Astronomy and Astrophysics, 21, 319

  \bibitem[Zheng et al.(2018)]{Zheng+etal+2018} Zheng, J., Zhao, G., Wang, W., et al. 2018, Research in Astronomy and Astrophysics, 18, 147
  
\end{thebibliography}
\end{document}